\documentclass[%
 reprint,
amsmath,amssymb,
 aps,
]{revtex4-2}

\usepackage{graphicx}
\usepackage{dcolumn}
\usepackage{bm}
\usepackage{physics}
\usepackage{braket}
\usepackage{amsmath,amssymb,amstext,mathtools}

\usepackage[pdftex,pagebackref=false]{hyperref} 
\hypersetup{
    plainpages=false,       
    unicode=false,          
    pdftoolbar=true,        
    pdfmenubar=true,        
    pdffitwindow=false,     
    pdfstartview={FitH},    
    pdfnewwindow=true,      
    colorlinks=true,        
    linkcolor=blue,         
    citecolor=green,        
    filecolor=magenta,      
    urlcolor=cyan           
}

\begin{document}

\preprint{APS/123-QED}

\title{ Optical field characterization at the fundamental limit of spatial resolution with a trapped ion}
\author{Nikhil Kotibhaskar}
\author{Sainath Motlakunta}
\author{Anthony Vogliano}
\author{Lewis Hahn}
\author{Rajibul Islam}
\affiliation{%
 Institute for Quantum Computing and Department of Physics and Astronomy, University of Waterloo, Waterloo, ON, N2L 3G1, Canada
}%


\newcommand{\Yb}{\ensuremath{^{171}\mathrm{Yb}^{+}\hspace{0.25em}}}
\newcommand{\Shalf}{\ensuremath{^2\mathrm{S}_{1/2}\hspace{0.25em}}}
\newcommand{\Phalf}{\ensuremath{^2\mathrm{P}_{1/2}\hspace{0.25em}}}

\newcommand{\Sixj}[6]{ 
\begin{Bmatrix}
  #1 & #2 & #3 \\
  #4 & #5 & #6 
\end{Bmatrix}
}
\newcommand{\kb}[2]{\ket{#1}\bra{#2}}

\begin{abstract}
Optical systems capable of generating fields with sub-wavelength spatial features have become standard in science and engineering research and industry. 
Pertinent examples include atom- and ion-based quantum computers and optical lithography setups.
So far, no tools exist to generally characterize such fields - both intensity and polarization - at sub-wavelength length scales. 
We use a single trapped atomic ion, confined to approximately $ 40 \;\mathrm{nm} \times  40 \;\mathrm{nm} \times 180 \;\mathrm{nm}$ to sense a laser light field at a wavelength of 370 nm.
With its spatial extent smaller than the absorption cross-section of a resonant detector, the ion-sensor operates at the fundamental limit of spatial resolution.
Our technique relies on developing an analytical model of the ion-light interaction and using the model to extract the intensity and polarization.
An important insight provided in this work is also that the inverse of this model can be learned, in a restricted sense, on a deep neural network, speeding up the intensity and polarization readout by five orders of magnitude. 
This speed-up makes the technique field-deployable to characterize optical instruments by probing light at the sub-wavelength scale.
\end{abstract}

\maketitle


\section{Introduction}
Sub-wavelength optical features are highly sought after for various applications in research and industry, including, but not limited to, nanofabrication \cite{Deubel2004DirectTelecommunications,TrangDo2013SubmicrometerWriting,Tsutsumi2017DirectPhotoresist}, high-resolution imaging \cite{Wilson1994ConfocalMicroscopy,Wang2016Scattering-typeSampling}, optical trapping \cite{Ashkin1987OpticalBeams}, high-capacity optical data storage \cite{Parthenopoulos1989Three-DimensionalMemory}, etc. 
Characterizing the size, shape, and polarization of such optical features is essential; however, creating compact and high-performance optical sensors for extracting light intensity and polarization presents considerable challenges, especially for polarization readout.
Over the past decades, significant advances have been made towards miniaturizing polarization sensitive devices, for example,  CCD-based polarization imaging devices \cite{Gruev2010CCDFilters}, plasmonic on-chip polarimeters \cite{Afshinmanesh2012MeasurementPolarimeter}, chiral beamsplitters based on gyroid photonic crystals \cite{Turner2013MiniatureCrystals}, compact polarization sensors and imagers using metasurfaces \cite{Lin2016EffectsIons,Wen2015MetasurfaceLight,Rubin2019MatrixCamera}.
However, the characterization of sub-wavelength optical  features produced in systems such as the atom- and ion-based quantum computers and optical lithography instruments would require resolution not yet available with the current devices.

The resolution of a sensing technique is ultimately limited by the spatial extent of the light probe or its absorption cross section \cite{Foot2004AtomicPress}, whichever is greater.
Typically, most detectors have a spatial extent larger than the absorption cross-section, and hence the detector size is generally the factor determining the resolution.
However, in the case of atomic and ionic systems, the spatial extent of their wavepackets can be made much smaller than the wavelength of optical transitions between their energy states.
For example, single trapped ions can be laser-cooled to be confined within $\mathcal{O}$(10 nm) \cite{Wong-Campos2016High-resolutionAtom}, much smaller than the typical ultraviolet or visible transition wavelengths, and hence the resolution is determined by the absorption cross-section.
The sub-wavelength confinement makes ions and atoms in optical tweezers \cite{Tomita2024AtomAtom} an excellent choice for high-resolution optical characterization.
Furthermore, the deep electronic trapping potential—without the need for complex optical trapping setups—combined with negligible perturbation from the light field to be characterized, and the high repetition rate of experiments make ions optimal for sub-wavelength sensing.

In this work, we demonstrate sub wavelength characterization of (UV) optical fields - both intensity and polarization- using a single trapped atomic ion as a sensor.
The ion is confined to a spatial extent of approximately  $ 40 \;\mathrm{nm} \times  40 \;\mathrm{nm} \times 180 \;\mathrm{nm}$, smaller than the fundamental cross-section for absorption by a resonant detector, $\lambda^2/2 \pi$ (in this work, $\lambda$ = 370 nm, i.e a circular cross-section of 185 nm diameter), and hence the resolution of our measurement is at the fundamental limit of spatial resolution.
To the best of our knowledge, this is the highest spatial-resolution optical characterization (intensity and polarization) to date.

\begin{figure*}
    \includegraphics[width=\linewidth]{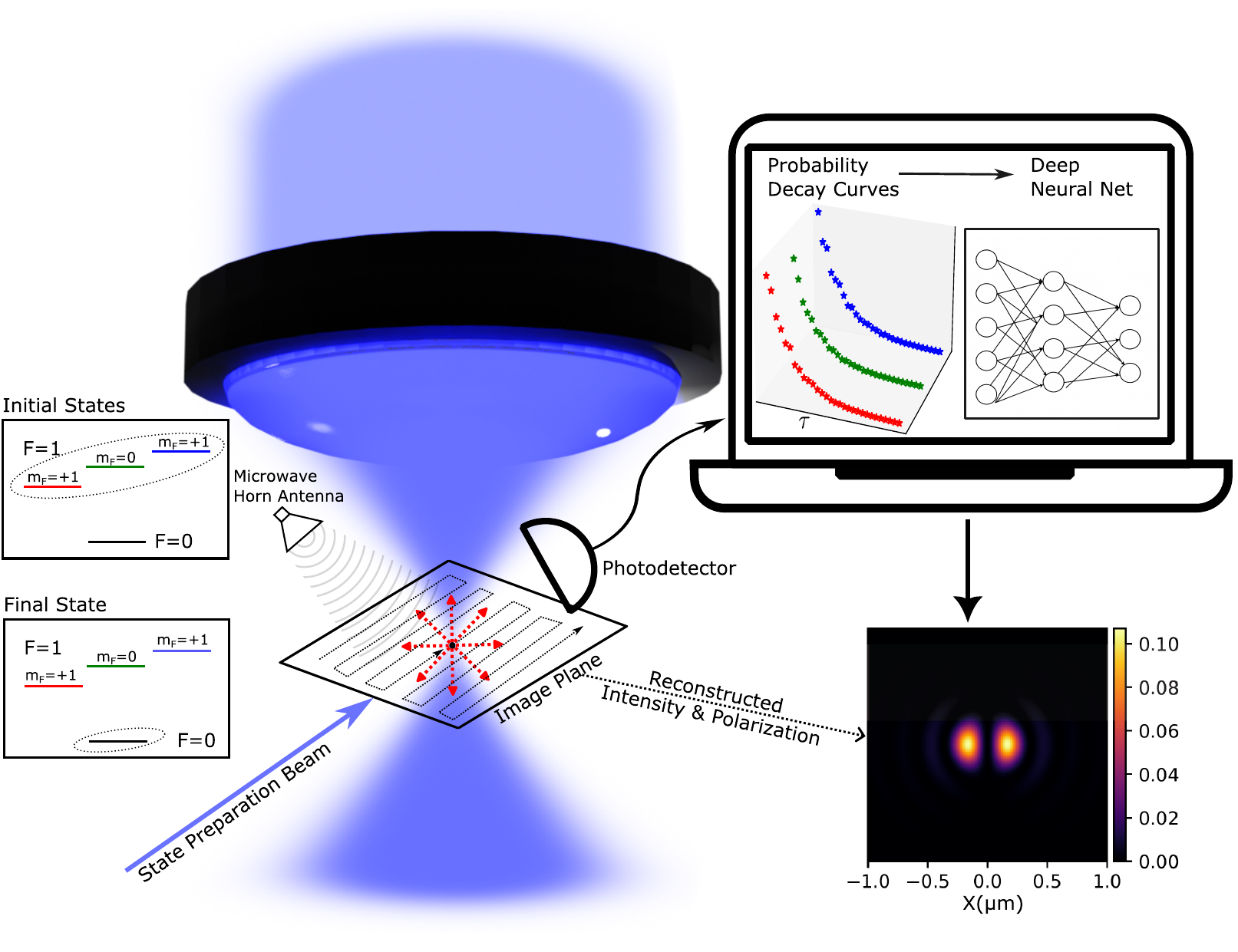}
    \caption{\textbf{Cartoon schematic of the light-field sensing technique:} A trapped \Yb ion is moved to the field point where the light properties are to be measured and initialized in one of the sublevels of the \Shalf $\ket{F=1}$ manifold (inset: Initial States) using a microwave horn antenna.
    The light that needs to be characterized pumps the population in \Shalf $\ket{F=1}$ manifold to the \Shalf $\ket{F=0}$ state (inset: Final State).
    The probability of pumping to \Shalf $\ket{F=0}$ is a function of the initial state of the ion, the interaction time $\tau$, and the light intensity and polarization.
    The three measured probability decay curves, with the ion initialized in each of the three initial states, are fed to a deep neural network which then outputs the intensity and polarization of the light at the ion position.} 
    \label{fig:Schematic}
\end{figure*}

Trapped ions are now a mature experimental platform and there are well-known protocols for the state initialization of ions, into particular quantum states, using optical pumping techniques \cite{Olmschenk2009QuantumQubits,Olmschenk2007ManipulationQubit,Monroe2021ProgrammableIons}).
The time evolution during the optical pumping process depends on the initial quantum state, as well as on the intensity and polarization of the light.
Our technique uses a well-characterized model of light-matter interaction, with no free parameters, to extract intensity and polarization from optical pumping time-series data.
We first experimentally validate our model, which includes the quantum evolution of eight atomic states involved in the process.
As a proof-of-principle demonstration, we measure the spatial profile of a tightly focused beam (waist $\approx$ 2.2 $\mu$m, $\lambda$ = 370 nm) with a resolution of 200 nm. 
To demonstrate the power of our protocol, we characterize the field at the focus of a high-resolution objective (NA = 0.8) beyond the capability of our hardware using synthetic data. 
Such high-resolution optical systems exhibit sub-wavelength features in the intensity and polarization of light that are inaccessible to existing optical characterization techniques.
Further, we adopt an ‘intelligent’ approach (Figure \ref{fig:Schematic}) that utilizes deep learning to accelerate, by more than $10^5$ times, the process of extracting optical parameters enabling characterization of the optical field with fine spatial resolution over a large region within a practically feasible time scale.

\section{Experimental Apparatus}
\label{sec:Apparatus}
\begin{figure*}
    \centering
    \includegraphics[width=\linewidth]{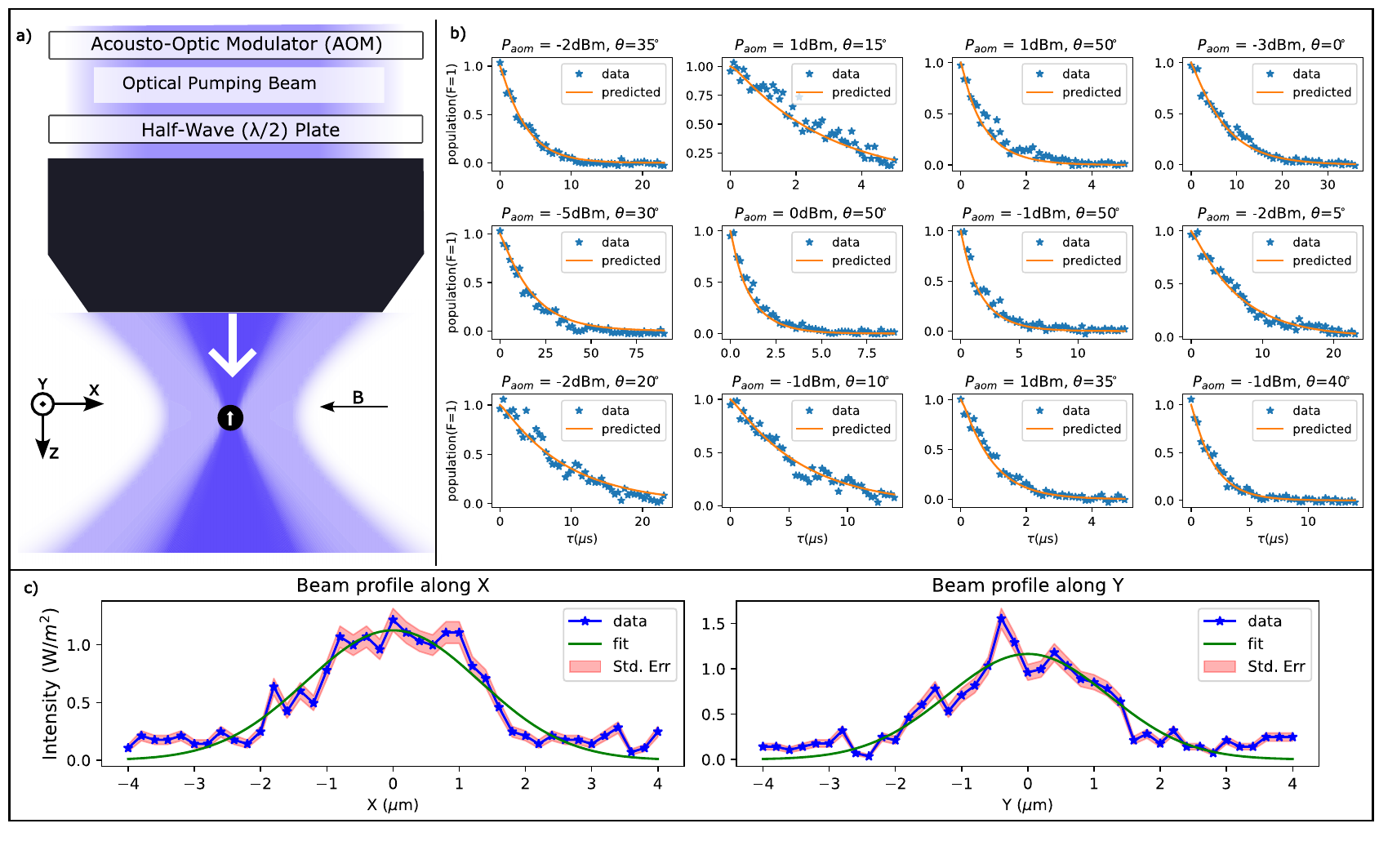}
    \caption{\textbf{Validating the ion-light interaction model:} 
    \textbf{a)} Cartoon schematic of the experimental apparatus used to generate the 88 experimental optical pumping curves for validating our model (See section \ref{sec:IonLightModelValidation}).
    76 curves were used to fit the data to the theoretical model and extracting the light parameters.
    \textbf{b)} Remaining 12 curves optical curves (not used for fitting) used for comparison with model predictions after extraction of light parameters. 
    Here, $P_{\mathrm{aom}}$ is the RF drive power to the AOM and $\theta$ is the half-wave plate angle.
    \textbf{c)} Measured beam profile of the characterization beam in steps of 200 nm.
    The shaded regions represent the standard errors (200 experimental repetitions) and the green line is a Gaussian fit. 
    The intensity at each point was estimated by recording probability of decay of the ion initialized in \Shalf $\ket{F=1, \;m_F=0}$ to \Shalf $\ket{F=0}$ after a fixed interaction time of $\tau = 30 \mu$s.
    }
    \label{fig:SolverValidate}
\end{figure*}
The experiments to validate our ion-light interaction model are carried out using a single \Yb ion confined in an ion trap with three harmonic oscillator frequencies $(\omega_x,\omega_y,\omega_z)/ 2 \pi$  = (1.1, 1.06, 0.27) MHz.
In the presence of a repump laser \cite{Olmschenk2007ManipulationQubit}, the \Shalf $\rightarrow$ \Phalf transition (D1) forms a closed cycle and the population can be expected to be confined to the \Shalf and \Phalf levels.
By choosing appropriate intensity, polarization, and frequency detuning from the the D1 transition (370 nm), and frequency modulation of light, we can initialize the ion in the \Shalf ground state through optical pumping, and perform fluorescence-based state readout \cite{Olmschenk2007ManipulationQubit}.
Figure \ref{fig:SolverValidate} a) shows a cartoon schematic of the apparatus used to generate the experimental optical pumping decay curves.
We use an objective lens to illuminate the ion with a well-focused beam (waist $\approx$ 2.2 $\mu$m, $\lambda$ = 370 nm), whose properties are to be measured, on the ion.
The same objective is used to collect the ion fluorescence, with a collection NA=0.16, to detect its state.
A magnetic field of 2.23 G is applied at the ion location to avoid inadvertent pumping to dark states during state detection \cite{Berkeland2002DestabilizationSystems}.
The intensity and polarization of the characterization beam can be varied by changing the radio frequency (RF) power driving an acousto-optic modulator (AOM) and the angle ($\theta$) of a wave plate, respectively.
The transfer function of the AOM, that is, RF power vs optical intensity (See supplementary for the detailed experimental apparatus), was characterized before the curves are recorded, and we use the standard theoretical transfer function for the wave plate. 

The laser-cooled ion has a position spread of $\Delta = \sqrt{2 \Bar{n} + 1} x_0$ where $x_0 = \sqrt{\hbar/2m\omega}$ is the zero-point spread of a quantum harmonic oscillator, $\Bar{n}$ is the mean thermal phonon occupation number and $\omega$ is the harmonic oscillator frequency (angular) of the ion in the trapping potential.
We estimate that at the doppler limit, the position spread of the ion is about 20 nm along the x and y directions, and 90 nm along z direction.
The size of the ion, seen by incoming light from any direction is smaller than the absorption cross-section of a resonant detector, i.e. $\lambda^2/2\pi$ \cite{Foot2004AtomicPress}, and the ion acts as a point detector only limited by this fundamental limit of spatial resolution.

We Doppler-cool the ion for 6 ms, followed by preparing the ion in the \Shalf $\ket{F=1, \;m_F=0}$ level, first by optical pumping \cite{Olmschenk2009QuantumQubits,Olmschenk2007ManipulationQubit,Monroe2021ProgrammableIons} to the state \Shalf $\ket{F=0}$, followed by a microwave transition (around 12.6 GHz) to \Shalf $\ket{F=1, \;m_F=0}$ (see Fig. \ref{fig:Schematic} Inset: Initial States).
The characterization beam is then incident on the ion for an interaction time $\tau$, and after this, the state of the ion is measured using the state-dependent fluorescence readout \cite{ActonM2006Near-perfectRegister,Hume2007High-fidelityMeasurements,Olmschenk2007ManipulationQubit}.
We repeat the experiment $N=200$ times to estimate the probability that the ion is optically pumped from the initial state in the \Shalf $\ket{F = 1}$ manifold to $\ket{F=0}$. 
We collect many optical pumping time-series datasets by varying $\tau$, with each dataset corresponding to a given initial state of the ion and a specific intensity and polarization of the characterization light.
We analyze these optical pumping curves to extract intensity and polarization.

\section{results}
\label{sec:results}

\subsection{Validating the ion-light model}
\label{sec:IonLightModelValidation}
We model the interaction of the ion with light with a semi-classical approach where we treat the light classically and the ion quantum mechanically.
We calculate the time evolution of the atomic density matrix corresponding to all energy levels in the ground (\Shalf) and excited states (\Phalf) in the presence of electromagnetic radiation. 
In order to efficiently solve for the time evolution of the atomic density matrix, the ion-light interaction Hamiltonian is converted to a time-independent Hamiltonian using a graph theory approach as is Ref. \cite{Einwohner1976AnalyticalApproximation} (See supplementary for details).  
The intensity of the light at the ion, $I$, is represented by the saturation parameter, $s=I/I_{\mathrm{sat}}$, where $I_{\mathrm{sat}}$ is the saturation intensity of the transition \cite{Foot2004AtomicPress}.
Polarization is specified with respect to the quantization axis set by the applied magnetic field, in the spherical bases, with the proportion of intensity in the left circular, right circular, and linear polarization denoted by the coefficients $(C_+, C_-, C_{\pi})$, respectively.

The propagation direction of the characterization beam, as shown in Fig. \ref{fig:SolverValidate} a), is perpendicular to the quantization axis \cite{Foot2004AtomicPress} (that is, the direction defining the polarization $\pi$ in the atomic frame).
Hence, for any angle of the wave plate $\theta$, the polarization of the beam remains in the XY plane.
We can write the Jones vector ($\theta = 0$) at the ion location as $J = (\epsilon_x, \epsilon_y + i\epsilon^\prime_y)$, and for a given light intensity these values can be converted to $(C_+, C_-, C_{\pi})$ as follows.
\begin{equation}
\label{eq:polarization}
\begin{split}
    C_{\pi} = |\epsilon_x \cos{(2\theta)} + (\epsilon_y + i\epsilon^\prime_y) \sin{(2\theta)}|^2 \\
    C_+ = C_- = |\epsilon_x \sin{(2\theta)} - (\epsilon_y + i\epsilon^\prime_y) \cos{(2\theta)}|^2 / 2
\end{split}
\end{equation}
Note that this neglects any component of the electric field in the beam propagation direction and this is valid since we are using a relatively low NA lens to focus the beam.

A total of 88 experimental curves are obtained, of which 76 are used for fitting (without any free parameters) and the remaining 12 for validation.
For fitting the experimental optical pumping curves to the model we start with an initial guess (generated by a reduced ion-light model, see methods) and vary $s, \epsilon_x, \epsilon_y, \epsilon^\prime_y$ to simultaneously minimize the least squares error between all the experimental curves and the ones predicted by the ion-light interaction model.
While fitting the optical pumping curves, we account for the measured scaling of the light intensity with respect to RF power input to the AOM, and the polarization variation at each waveplate angle.
After minimization, we find $s = 0.44662(7)$ $\epsilon_x = 0.8260(2)$, $\epsilon_y = 0.38595(8)$, $\epsilon^\prime_y = 0.4107(1)$ for $\theta = 0$ and an AOM RF power of 3 dBm.
The polarization and intensity for the other values of $\theta$ and AOM RF power can be found using Eq. (\ref{eq:polarization}) and the measured scaling of the light intensity with respect to RF power.
The fitting procedure takes $\approx$ 1 hour to perform on an 8 core CPU (AMD Ryzen 7 2700x).
We further corroborate the validity of the model by generating predictions for the remaining 12 experimental optical pumping curves after accounting for the AOM power and waveplate angle.
Figure \ref{fig:SolverValidate} b) shows these 12 curves and the corresponding predictions from the model.

It is important to mention that the chosen data set includes more curves than necessary to determine light intensity and polarization. 
This highly constrained data set (due to the AOM and waveplate transfer functions) is intended to verify the simulation model rather than extract light parameters, which is an incidental result. 
The subsequent section discusses a more effective algorithm specifically for estimating light parameters.

To demonstrate the high spatial resolution of the technique, we measured the beam profile of the characterization beam by moving it with respect to the ion in increments of 200 nm using our previously demonstrated method for precise beam profile generation using Fourier holography \cite{ShihChung-You2021ReprogrammableControl}.
High NA lenses have a polarization gradient in the image plane; however, for our relatively low NA no such variation is expected, simplifying the beam profile measurement.
Given that we have already extracted the beam polarization, we can measure the beam profile by performing only a single optical pumping measurement, with a fixed interaction time $\tau$ = 30 $\mu$s, each time the characterization beam was moved with respect to the ion.
Figure \ref{fig:SolverValidate} \textbf{c)} shows the measured beam profile of the characterization beam along the X and Y directions.
For high NA systems, where the polarization gradient is significant, this technique can be used to measure the polarization gradient in the image plane.

\subsection{Intelligent Sensing}
\label{sec:IntelligentSensing}

\begin{figure*}
    \centering
    \includegraphics[width=0.8\linewidth]{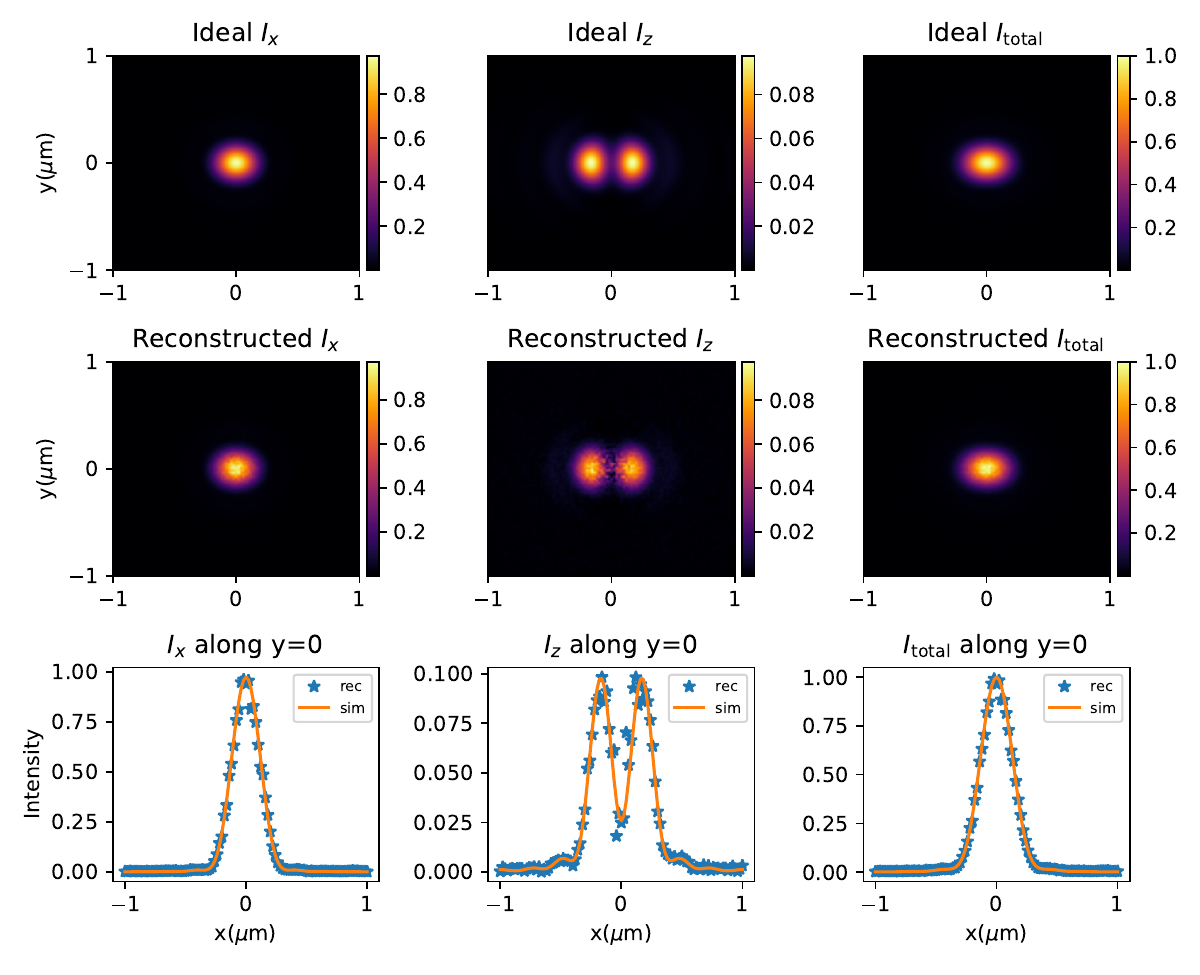}
    \caption{\textbf{Intelligent Sensing:}    
    \textbf{Top}: Simulated focal spot of a 0.8 NA objective lens assuming perfect reconstruction with a detector of cross-section $\lambda^2/2\pi$, at the wavelength $\lambda = 370$ nm.
    Here, just like our apparatus in Fig. \ref{fig:SolverValidate}, the z-axis is the direction of propagation of the beam at the input of the objective lens.
    The input light is x-polarized and $I_\mathrm{x}$, $I_\mathrm{z}$ and $I_{\mathrm{total}}$ represent normalized intensities.
    \textbf{Middle}: Reconstructed image plane field using intelligent reconstruction.
    \textbf{Bottom}: Reconstructed image plane field using intelligent reconstruction: Slice along y=0.
    }
    \label{fig:highNAreconstruction}
\end{figure*}

Using high-NA objective lenses, tightly focused spots with sub-wavelength dimensions can be achieved.
Such features have both intensity and polarization variation at the sub-wavelength scale.
Characterizing the optical performance of such instruments is a major application of this technique, but the computational cost of fitting is the primary overhead.
For example, measuring the field in a 1 mm $\times$ 1 mm region with a resolution of 200 nm requires spatial locations $\mathcal{O}(10^7)$.
To characterize the optical field in the image plane of a high NA optical instrument, a more efficient reconstruction algorithm is required.

One way to reduce the computational overhead in light parameter estimation is to find the minimum number of optical pumping curves required to extract light parameters at each field point.
For a given initial state of the ion, multiple values of intensity and polarization can lead to very similar curves, and hence one optical pumping decay curve is not enough to extract the light characteristics.
However, the internal energy structure of the \Yb ion and the atomic selection rules allow us to devise a strategy to estimate the intensity and polarization.
We find that initializing the ion in each of the three sublevels ($m_F = -1,0,1$) of the \Shalf $F=1$ manifold and recording the optical pumping curve for each state makes it possible to reconstruct the light parameters.
This makes intuitive sense since each of these initial states is unperturbed by exactly one of the three polarizations $\sigma_+, \sigma_-, \pi$ \cite{Olmschenk2007ManipulationQubit}.
Parameter extraction using this method requires $\mathcal{O}$(1 minute) per field point on an 8-core CPU (AMD Ryzen 7 2700x), with the computational cost being dominated by calls to the ion-light interaction model during the fitting process.

To completely bypass the fitting process, we aim to solve the inverse problem \cite{NewtonInversePhysics} that corresponds to finding the functional inverse of the model that describes the ion-light interaction.
It turns out that the specific function $F_\mathrm{fwd}$, which takes the light parameters as input, and provides as output the three separate optical pumping curves for an ion initialized in the $m_F = -1,0,1$ states of the \Shalf F=1 manifold respectively, can be inverted using a deep neural network.
After finding the inverse function, $F_\mathrm{inv}$, we can input the three experimental optical pumping curves, one each for the ion initialized in the $m_F = -1,0,1$ states, and it outputs the light parameters $s, (C_+, C_-, C_{\pi})$.
The use of a deep neural network to approximate $F_\mathrm{inv}$ reduces the reconstruction time to $\mathcal{O}(66 \; \mu \mathrm{s})$ per field point on an 8-core CPU (AMD Ryzen 7 2700x).

As an application of this technique, we wanted to characterize the image plane field of a High-NA optical instrument. 
In our current experimental setup, we are limited to a relatively low NA (0.16) focusing lens; however, simulating the image plane field of a high-NA lens is relatively straightforward, and hence we test the efficacy of our technique to characterize such a device using synthetic data.

The focal spots or point spread functions of high NA optical objective lenses have been theoretically studied in detail in the literature \cite{Woehl2010RealisticMicroscopy,Ganic2003FocusingSpace,Li2015InfluenceSpot}, and much of the modeling follows the approach of the vector Debye model proposed by Richards and Wolf \cite{Richards1959ElectromagneticSystem}.
Figure \ref{fig:highNAreconstruction} (Top) shows the simulated focal spot of a lens with NA = 0.8 measured on an ideal resonant detector on a 1 $\mu$m $\times$ 1 $\mu$m grid.
To account for the detector's fundamental cross-section, the point spread function calculated from the vector Debye model was blurred using a Gaussian filter with a kernel size approximately equal to the cross-section diameter ($\approx$ 185 nm) of a resonant detector at $\lambda$ = 370 nm.

Since we have access to our ion-light simulation model and have already corroborated its validity, we can use it to simulate the ion's optical pumping curves at each field point (1 $\mu$m $\times$ 1 $\mu$m grid with a 20 nm step).
We generate three optical pumping curves, one each for the ion initialized in the $m_F = -1,0,1$ states of the \Shalf F=1 manifold using the light parameters at each field point.
Although the state preparation and measurement fidelity of state-of-the-art ion trap quantum devices are above 99.9\%; to mimic a highly noisy experimental scenario, we added a noise of 10 \%. 
Adding noise here refers to errors in the estimation of the population at each point in an optical pumping curve.
For example, an error of 10\% means that at each point in an optical pumping curve, the population measured in the F = 1 manifold can be off by $\pm$ 5\%.
These noisy optical pumping curves were then directly inputted into the deep neural network ($F_\mathrm{inv}$) and Fig. \ref{fig:highNAreconstruction} (middle and bottom) shows the performance of the neural network in reconstructing the image plane field using the noisy optical pumping curves.
Even with such a high noise, the reconstruction
performance points towards low experimental requirements of this method.

\section{Outlook}
\label{sec:Outlook}
We have presented a method for characterizing optical instruments producing sub-wavelength optical features using a single ion as a light sensor.
This technique is underpinned by our novel approach to modeling the ion-light interaction, which in itself offers a valuable tool for atomic physics laboratories seeking to characterize their experimental setups.
The resolution of the current technique can be further improved by making use of off-resonant detection, since the absorption cross-section reduces as the laser frequency is tuned away from resonance.
For example, for the \Shalf to \Phalf transition in \Yb (linewidth $\Gamma \approx$ 20 MHz) used in this work, tuning the laser out of resonance by $\Gamma$ results in a cross-section reduction by a factor of 5 \cite{Foot2004AtomicPress} bringing the detector size below 40 nm.

A further reduction in the cross-section can be achieved by measuring intensity- and polarization-dependent stark shifts of the internal energy levels \cite{Degen2017QuantumSensing} using Ramsey interferometry \cite{Ramsey1950AFields}.
In \Yb this would translate into measuring the Stark shifts on the \Shalf, $m_f = 1,0,-1$ levels, induced by an off-resonant beam on the \Shalf to \Phalf transition. 
Several apparatuses using the \Yb ion (including the one used in this work) use the 355 nm lasers off resonantly couple these levels, and this could be a readily usable wavelength to characterize optical instruments.
The spatial extent of the ions, with appropriate cooling technique, would then be the only limiting factor for the resolution, and, as discussed earlier, this can be made $\mathcal{O}$(10\;nm).

The wide adoption of this technique for characterizing optical systems would require the overcoming of two main challenges.
The first challenge is that the trapped ions are inside a vacuum system and that the optical field that needs to be characterized should pass through a vacuum window (effectively a thin slab of glass) before reaching the ions.
The vacuum window limits the minimum working distance and, more importantly, causes distortions to the optical field.
The transfer function of the window should therefore be accounted for using independent characterization or modeling.
For the special case of ion-based quantum information processing devices, the effect of the vacuum window is usually accounted for while designing the microscope objective itself.
Therefore, our technique can be readily applied to characterize qubit-addressing light in such devices.
Another option would be to custom-build a vacuum system that houses both the ion trap and the objective under test. 
That way, the light could pass the vacuum viewport when it is a collimated beam, leading to minimal distortion of the field.
Many systems in nanofabrication laboratories and cleanrooms, for example, sputtering and thermal deposition chambers, have vacuum systems that enable installation of samples and pumping to ultrahigh vacuum.

The second challenge stems from the requirement that the light must be resonant with certain transitions of the ions and this seems to limit the applicability of this technique for characterization of optical systems for use with multiple wavelengths or white light.
Each ion species has several transitions that can be used for sensing; for example, the \Yb ion has transitions at 369 nm, 411 nm, 435 nm, and 935 nm that could be used for sensing.
Moreover, multispecies ion traps, i.e. traps that can confine different species of ions at the same time, are very common and this can further increase the optical bandwidth of the system.
Since only a few wavelengths are required to characterize an imaging system, the resonance requirement of this technique does not pose any fundamental limitation in the characterization of multiwavelength optical systems.
Similarly, the Stark shift-based approach \cite{Tomita2024AtomAtom} does not require resonant detection and, hence, does not suffer from this limitation.

The light field characterization technique presented in this work opens new avenues for the characterization of complex optical systems capable of producing sub-wavelength optical features with unprecedented resolution.
This is especially useful in characterizing light sources in optical lithography systems and atom or ion-based quantum computers.
The intelligent characterization technique will also speed up the calibration of atom or ion-based quantum computers, increasing their computing efficiency.

\section{Methods}
\label{sec:Methods}

\subsection{Modelling atom-light Interactions}
The full solver takes into account all 8 energy levels in the \Shalf (ground) and the \Phalf (excited) manifold.
Although only optical pumping light was used in this work, the full solver accounts for the effect of all the relevant light and microwave frequencies such as the cooling , state detection and spin flipping microwave transition coupling the clock states in the \Shalf manifold.  
The Hamiltonian that describes the interaction of ions with light after a standard rotating wave approximation \cite{Foot2004AtomicPress} can be written as 

\begin{equation}
    H_{int} = \sum_i \omega_i \kb{i}{i} + \sum_{i,j,l} \frac{\Omega_{ij}}{2} \left(\kb{i}{j} \times e^{i\omega_l t} + \text{h.c.} \right)
\end{equation}
Where $i,j$ are the indices that span the 8 energy levels, the index $l$ accounts for all light frequencies ($\omega_l$).
The above Hamiltonian is simplified to a time-independent Hamiltonian using a graph-theoretic approach \cite{Einwohner1976AnalyticalApproximation} to yield as simplified hamiltonin as follows,

\begin{equation} \label{eq:timeIndependentHamiltonian}
    H_{int}^{\prime} = \sum_i \Delta_i \kb{i}{i} + \sum_{i,j} \frac{\Omega_{ij}}{2} \left( \kb{i}{j} + \kb{j}{i} \right).
\end{equation}

The Rabi frequencies, $\Omega_{ij}$, for the coupling states $\ket{i}$ and $\ket{j}$ are calculated by evaluating the relevant coupling matrix elements according to the following expression (see supplementary for full derivation)
\begin{equation}
    \Omega^2_{ij} = \frac{I_{ij}}{I_{\mathrm{sat}}} \frac{\Gamma^2}{6}.
\end{equation}
Where $I_{ij}$ is the intensity of the light component that couples the levels $i$ and $j$.
To account for the spontaneous decay of the \Phalf manifold, we write the Lindblad (or Gorini-Kossakowski-Sudarshan-Lindblad) master equation with the relevant collapse for spontaneous emission as per the atomic selection rules.
We then numerically calculate the evolution of the entire atomic density matrix, using qutip \cite{Johansson2013QuTiPSystems}.
The modeling is covered in detail in the supplementary section.
This work also serves as a reference for modeling the common operations on the \Yb ion (the method also works for other common ions), such as optical pumping, detection, cooling and spin flips, and an important insight is that these operations, and hence complicated sequences of them, can be efficiently simulated on a classical computer.

\textbf{A Reduced Model -} The second method, used primarily to generate fit guesses, utilizes a simplified method to model the optical pumping phenomenon in multilevel atoms \cite{Atoneche2017SimplifiedPumping}.
Here, each pair of ground and excited sublevels, connected by light, is considered an effective two-level system, and a differential equation is obtained for the population dynamics of the four levels in \Shalf.
If the radiative lifetime of any of the excited states $e_k$ is $\tau_k$, then the natural linewidth of any transition to the excited state will be 
\begin{equation}
    \Gamma_k = \frac{1}{\tau_k} = \sum_{j=1}^{N_g} A_{jk} = \Gamma_k \sum_{j=1}^{N_g} \beta_{jk}. 
\end{equation}
Here, $\beta_{jk}$ is the branching ratio of the radiative decay from state $e_k$ to $g_j$ and we have that $\sum_{j=1}^{N_g} \beta_{jk} = 1$.
The notation here is such that the first index denotes the levels in the ground manifold, and the second index denotes the levels in the excited manifold.
Since the idea here is to treat each 2-level system of coupled ground and excited states separately, we only require an expression for the rate of excitation from the $i^{th}$ ground state to the $j^{th}$ excited state $R_{jk}^l$.
From standard textbook calculation for two-level systems \cite{Foot2004AtomicPress} we have that
\begin{equation}
    R_{jk}^l(s,\Delta) = \frac{ \Gamma_k}{2} \frac{s}{1 + s + (2\Delta_{jk}^l)/\Gamma_k)^2},
\end{equation}

\begin{equation}
    \begin{split}
        \frac{d}{dt} G_n(t) &= \sum_{j \neq n} G_j(t) \left[ \sum_k R_{jk}^l \beta_{jk} \beta_{nk} \right] \\
         &- G_n(t) \left[ \sum_k R_{nk}^l\beta_{nk}(1-\beta_{nk}) \right].
    \end{split}
\end{equation}

This differential equation can be solved numerically to obtain the ion's characteristic population decay curves in the presence of the characterization light.
This reduced model only simulates the population dynamics in the \Shalf manifold and does not evolve the full 8 level density matrix, and all details of the model can be found in the Supplementary Document.
We find that the two models produce results consistent with each other upto $s$ = 0.5 after which the predictions start to diverge.
This is expected since the approximations in the generalized optical pumping model work best at low intensities.
The rationale behind developing the reduced (second) model was not only to serve as a sanity check, but also to create a simulation model with a lower computational cost.
This is useful for generating good initial guesses for the light intensity and polarization when fitting the experimental data to the full model.

\subsection{Deep neural network for Intelligent Sensing}
As mentionaed in the text the function $F^\prime_{\mathrm{inv}}$ was approximated using a deep neural network.
Note that this is the inverse of the specifically chosen function, $F_\mathrm{fwd}$, which takes the light parameters as input, and provides as output the three separate optical pumping curves for an ion initialized in the $m_F = -1,0,1$ states of the \Shalf F=1 manifold.
To train the neural network, a large number of solutions of $F_\mathrm{fwd}$ with known input values was simulated.
The parameter values for these simulations were chosen such that the $s$ parameter for each component of optical pumping light, 
\begin{equation}
    s_i = s \times C_i \; \text{for} \; i \in \{ +, -, \pi \},
\end{equation}
contained 101 evenly spaced values in log scale from $10^{-4}$ to 1.
The simulation time for each simulation was chosen to be 20 microseconds.
Since each of these simulations is independent of each other and hence can be parallelized, leading to large savings in simulation time, and in this case, this was approximately a 16X speedup due to the 16-core CPU used.
The entire training set took $\approx$ 6 hours to simulate.

The structure of the general network is an input layer, followed by four hidden layers and an output layer.
The input of the neural network is three optical pumping curves with 100 points each stacked to form a single vector, leading to an input layer size of 300 neurons.
The number of neurons in the four hidden layers is 2400, 600, 120, and 20, respectively, and the output layer has three neurons.
The activation function for all neurons in the hidden layers is a rectified linear unit (ReLU), and all layers, including the input, have full connectivity, i.e., each neuron in a given layer is connected to all the neurons in the next layer.
For training the network, the initial data set with $101^3$ samples was divided into \textit{training}, \textit{validation} and \textit{test} datasets with 90\% 9\% and 1\% of the samples in each data set, respectively.
To optimize the parameters of the neural network, the ADAM algorithm \cite{Kingma2014Adam:Optimization} was used to minimize, over all samples in the training set, the following mean absolute error loss function. 

\begin{equation}
    L = \sum_n^{N^\prime} \sqrt{(s_+^n - S_+^n)^2 + (s_-^n - S_-^n)^2 + (s_\pi^n - S_\pi^n)^2},
\end{equation}
where $N^\prime$ is the number of samples in the data set, ($s_+^n, s_-^n, s_\pi^n$) are the s parameters for the three polarization components of the optical pumping light and ($S_+^n, S_-^n, S_\pi^n$) are the values predicted by the neural network.

\begin{acknowledgments}
We acknowledge Irene Melgarejo Lermas for helpful discussions on deep neural networks.
We acknowledge financial support from the Canada First Research Excellence Fund (CFREF), the Natural Sciences and Engineering Research Council of Canada (NSERC) Discovery program (RGPIN-2018-05250), the Government of Canada’s New
Frontiers in Research Fund (NFRF), Ontario Early Researcher Award, University of Waterloo, and Innovation, Science and Economic Development Canada (ISED).
\end{acknowledgments}

\bibliography{references}

\pagebreak
\onecolumngrid
\section{Supplementary}
\subsection{Experimental Setup: Detailed Schematic}
\begin{figure*}[h]
    \centering
    \includegraphics[width=0.8\linewidth]{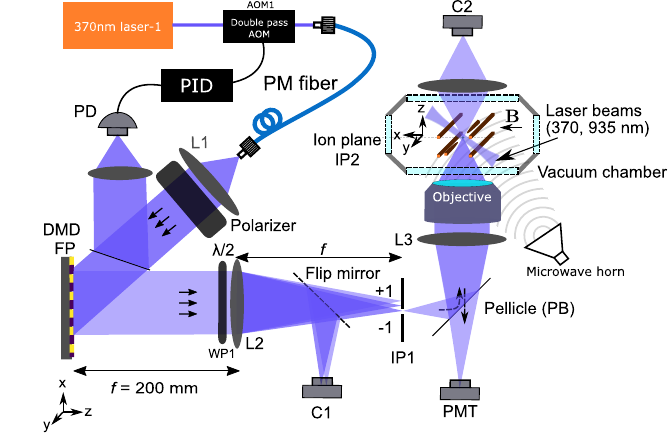}
    \caption{Detailed Schematic of the experimental setup}
    \label{fig:SuppsensorSchematic}

\end{figure*}

Figure \ref{fig:SuppsensorSchematic} shows the detailed schematic of the experimental setup used in this work.
The state preparation, detection and cooling light (370 nm,935 nm) are incident on the ions from a  direction perpendicular to the probe beam (370 nm).
THe 935 nm beam is the repump beam that allows the ion population to be confined to the \Shalf and the \Phalf manifolds.
The microwave transitions described in the text are achieved by impinging the ion with microwave radiation (12.642813 GHz) using a microwave horn.  
The probe beam at 370 nm is derived from a separate laser (370nm-laser-1) from the one that generates the state preparation, detection and cooling light (370nm-laser-2).

The probe beam, generated by the 370nm-laser-1, is focussed onto the ions using the following addressing setup that allows for precise scanning of the beam over the ions \cite{ShihChung-You2021ReprogrammableControl}.
An acoustic-optic modulator (AOM1) in a double pass configuration, placed after the 370nm-laser-1, is used as a switch with precise timing and power control for the probe light.
The light is then coupled to a PM fiber which is then expanded using a single lens(L1) and is polarization-cleaned using a polarizer. 
The light is sampled onto a photodiode (PD) that is used to stabilize the intensity fluctuations using PID feedback to the AOM.
The polarization-cleaned and power-stabilized light from the PM fiber illuminates a Digital Micromirror device (DMD) (Visitech Luxbeam 4600 DLP) placed in the Fourier plane. 
A motorized $\lambda /2$ waveplate(WP1) is placed after the DMD to control the final polarization of the light.
The negative first-order beam diffracted from the hologram on DMD is then relayed to the ion through the reflection of the pellicle beam splitter(45:55)(Thorlabs: BP145B5).
The pellicle beam splitter also allows the fluorescence from the ion, collected for state detection using the same objective for addressing the ion, to be collected at the PMT(Hamamatsu:H10682-210). 
A flip mirror placed before the intermediate image plane IP1 is used to image the IP1 onto a camera C1 for initial characterization.
The camera C1 is also used to calibrate the AOM power vs the optical intensity of the probe beam.
Since we use an optical fiber for the mode cleanup, the beam profile does not change after the fiber output when the RF power of the AOM is varied and hence measuring the scaling of the optical power at C1 w.r.t the RF power is sufficient to calibate the intensity scaling.
Figure shows the recored intensity scaling of the optical power at C1 with respect to the RF power.

\subsection{Ion-light interactions: Full solver}
\begin{figure}[h]
    \centering
    \includegraphics[width=0.6\linewidth]{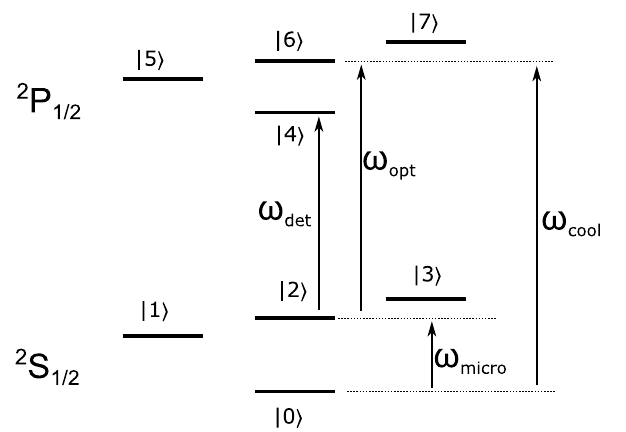}
    \caption{Labeling Scheme for energy levels in \Shalf  and \Phalf}
    \label{fig:YBEnergyLevels}
    With the above labeling convention, $\omega_{\mathrm{hf}}^\mathrm{S}$ is the frequency difference between $\ket{0}$ and $\ket{2}$,  $\omega_{\mathrm{hf}}^\mathrm{P}$ is the frequency difference  between $\ket{4}$ and $\ket{6}$, $\omega_{\mathrm{zm}}^\mathrm{S}$ is the frequency difference between $\ket{1}$ and $\ket{2}$, and $\omega_{\mathrm{zm}}^\mathrm{P}$ is the frequency difference between $\ket{6}$ and $\ket{7}$
    
\end{figure}

To model the dynamics in the 8-level subspace of \Shalf and \Phalf, we can start by labeling the energy levels as shown in Fig. \ref{fig:YBEnergyLevels}.
With $\ket{i}$ as the basis ket for level $i$, we can define the atomic Hamiltonian as follows

\begin{equation}
    \begin{split}
        H_{\mathrm{atom}} &= \kb{1}{1}  \omega_{\mathrm{hf}}^\mathrm{S} - \omega_{\mathrm{zm}}^\mathrm{S} 
        + \kb{2}{2}  \omega_{\mathrm{hf}}^\mathrm{S}
        + \kb{3}{3}  \omega_{\mathrm{hf}}^\mathrm{S} + \omega_{\mathrm{zm}}^\mathrm{S} \\ &
        + \kb{4}{4}  \omega_{\mathrm{SP}} + \omega_{\mathrm{hf}}^\mathrm{S} 
        + \kb{5}{5}  \omega_{\mathrm{SP}} + \omega_{\mathrm{hf}}^\mathrm{P} + \omega_{\mathrm{hf}}^\mathrm{S} - \omega_{\mathrm{zm}}^\mathrm{P} \\ &
        + \kb{6}{6}  \omega_{\mathrm{SP}} + \omega_{\mathrm{hf}}^\mathrm{P} + \omega_{\mathrm{hf}}^\mathrm{S}
        + \kb{7}{7}  \omega_{\mathrm{SP}} + \omega_{\mathrm{hf}}^\mathrm{P} + \omega_{\mathrm{hf}}^\mathrm{S} + \omega_{\mathrm{zm}}^\mathrm{P}
    \end{split}
\end{equation}

Here, $\omega_{\mathrm{SP}}$ is the frequency difference between the \Shalf and \Phalf levels, $\omega_{\mathrm{hf}}^\mathrm{S}$ is the hyperfine splitting in the \Shalf, $\omega_{\mathrm{hf}}^\mathrm{P}$ is the hyperfine splitting in \Phalf, $\omega_{\mathrm{zm}}^\mathrm{S}$ is the Zeeman splitting in \Shalf, and $\omega_{\mathrm{zm}}^\mathrm{P}$ is the Zeeman splitting in \Phalf.

We now model the interactions with light and microwave radiation at various frequencies.
If we ignore the Zeeman splitting in \Shalf and \Phalf, there are three optical transitions possible, i.e. \Shalf F = 0 $\rightarrow$ \Phalf F$^\prime$ = 1, \Shalf F = 1 $\rightarrow$ \Phalf F$^\prime$ = 0, \Shalf F = 1 $\rightarrow$ \Phalf F$^\prime$ = 1.
Since the Zeeman splitting in our experiment is smaller than the linewidth of the \Shalf to \Phalf transition, only one frequency is considered for each of these transitions, and the allowed transitions between the Zeeman levels are induced by the same.
The Hamiltonian for the light-ion interaction is a straightforward dipole interaction Hamiltonian of the form.

\begin{equation}
    H_{light} = - \Vec{d} \cdot \Vec{E}(t)
\end{equation}
Where $\Vec{E}(t) = \Vec{E}_\mathrm{opt} \cos(\omega_{\mathrm{opt}} t) + \Vec{E}_\mathrm{det} \cos(\omega_{\mathrm{det}} t) + \Vec{E}_\mathrm{cool} \cos(\omega_{\mathrm{cool}} t)$.
Here, the three frequencies $\omega_{\mathrm{opt}}$, $\omega_{\mathrm{det}}$, and $\omega_{\mathrm{cool}}$ correspond to the transitions F=1 $\rightarrow$ F $^\prime$ = 1, F=1 $\rightarrow$ F $^\prime$ = 0, and F=0 $\rightarrow$ F $^\prime$ = 1 respectively.

\begin{equation}
    \begin{split}
        H_{\mathrm{light}} &= \frac{1}{2} \bigg( 
            \Omega_{15} \kb{1}{5} + \Omega_{16} \kb{1}{6} + \Omega_{25} \kb{2}{5} + \cdots \\
            &+ \Omega_{27} \kb{2}{7} + \Omega_{36} \kb{3}{6} + \kb{3}{7} 
        \bigg) e^{i\omega_{\mathrm{opt}} t } \\
        &+ \frac{1}{2} \bigg( 
            \Omega_{14} \kb{1}{4} + \Omega_{24} \kb{2}{4} + \Omega_{34} \kb{3}{4} 
        \bigg) e^{i\omega_{\mathrm{det}} t } \\
        &+ \frac{1}{2} \bigg( 
            \Omega_{05} \kb{0}{5} + \Omega_{06} \kb{0}{6} + \Omega_{07} \kb{0}{7} 
        \bigg) e^{i\omega_{\mathrm{cool}} t } + h.c.
    \end{split}
\end{equation}

The Rabi frequencies are defined as follows.

\begin{equation}
\Omega_{ij} = \left\{\begin{array}{ll}
    \bra{i} \Vec{d} \cdot \Vec{E}_\mathrm{opt}\ket{j}/\hbar & \text{ if } F=1 \rightarrow F^\prime = 1 \\
    \bra{i} \Vec{d} \cdot \Vec{E}_\mathrm{det}\ket{j}/\hbar & \text{ if } F=1 \rightarrow F^\prime = 0  \\
    \bra{i} \Vec{d} \cdot \Vec{E}_\mathrm{cool}\ket{j}/\hbar & \text{ if } F=0 \rightarrow F^\prime = 1
\end{array}\right.
\end{equation}
In this way, we are only considering the resonant couplings.

The only other applied field to consider is the microwave field applied to couple the clock states.

\begin{equation}
    H_{\mathrm{micro}} = \frac{1}{2}\bigg( \Omega_{02} \kb{0}{2} \bigg)e^{-\omega_{\mathrm{micro}} t} + h.c.
\end{equation}
The total Hamiltonian of the atom is then given as

\begin{equation} \label{eq:CH4totalH}
    H_{\mathrm{tot}} = H_{\mathrm{atom}} + H_{\mathrm{opt}} + H_{\mathrm{det}} + H_{\mathrm{cool}} + H_{\mathrm{micro}}
\end{equation}
To consider off-resonant interactions, we can define the detunings as follows.

\begin{equation}
\begin{aligned}
\Delta_{\mathrm{mic}} &= \omega_{\mathrm{micro}} - \omega_{\mathrm{hf}}^\mathrm{S} \\
\Delta_{\mathrm{opt}} &= \omega_{\mathrm{opt}} - \omega_{\mathrm{SP}} - \omega_{\mathrm{hf}}^\mathrm{P} \\
\Delta_{\mathrm{det}} &= \omega_{\mathrm{det}} - \omega_{\mathrm{SP}} \\
\Delta_{\mathrm{cool}} &= \omega_{\mathrm{cool}} - \omega_{\mathrm{hf}}^\mathrm{S} - \omega_{\mathrm{SP}} - \omega_{\mathrm{hf}}^\mathrm{P}
\end{aligned}
\end{equation}

With the above definitions, the total Hamiltonian in the interaction frame of the atom ($H_{\mathrm{rot}} = UHU^{\dagger} + i\hbar \dot U U^{\dagger}$ with $U = e^{iH_{\mathrm{atom}}t}$) is given as

\begin{equation}\label{eq:CH4Hrot}
\begin{aligned}
H_{\mathrm{rot}} =  \frac{1}{2}&\bigg(
\kb{0}{2} \Omega_{02} e^{i \Delta_{\mathrm{mic}} t} + \kb{0}{5} \Omega_{05} e^{i t (\Delta_{\mathrm{cool}} + \omega_{\mathrm{zm}}^\mathrm{P})} + \kb{0}{6} \Omega_{06} e^{i \Delta_{\mathrm{cool}} t} \\
&+ \kb{0}{7} \Omega_{07} e^{i t (\Delta_{\mathrm{cool}} - \omega_{\mathrm{zm}}^\mathrm{P})} + \kb{1}{4} \Omega_{14} e^{i t (\Delta_{\mathrm{det}} - \omega_{\mathrm{zm}}^\mathrm{S})} + \kb{1}{5} \Omega_{15} e^{i t (\Delta_{\mathrm{opt}} + \omega_{\mathrm{zm}}^\mathrm{P} - \omega_{\mathrm{zm}}^\mathrm{S})} \\
&+ \kb{1}{6} \Omega_{16} e^{i t (\Delta_{\mathrm{opt}} - \omega_{\mathrm{zm}}^\mathrm{S})} + \kb{2}{4} \Omega_{24} e^{i \Delta_{\mathrm{det}} t} + \kb{2}{5} \Omega_{25} e^{i t (\Delta_{\mathrm{opt}} + \omega_{\mathrm{zm}}^\mathrm{P})} \\
&+ \kb{2}{7} \Omega_{27} e^{i t (\Delta_{\mathrm{opt}} - \omega_{\mathrm{zm}}^\mathrm{P})} + \kb{3}{4} \Omega_{34} e^{i t (\Delta_{\mathrm{det}} + \omega_{\mathrm{zm}}^\mathrm{S})} + \kb{3}{6} \Omega_{36} e^{i t (\Delta_{\mathrm{opt}} + \omega_{\mathrm{zm}}^\mathrm{S})} \\
&+ \kb{3}{7} \Omega_{37} e^{i t (\Delta_{\mathrm{opt}} - \omega_{\mathrm{zm}}^\mathrm{P} + \omega_{\mathrm{zm}}^\mathrm{S})} \bigg) + h.c.
\end{aligned}
\end{equation}
The above Hamiltonian has time dependence at several frequencies, and since the goal of the modeling is to have an efficient numerical simulation of the ion-light dynamics, removing the time dependence of the Hamiltonian is essential.
In order to remove the time dependence, we need to switch into an interaction frame that removes the time dependence.
We can proceed with first defining a unitary transformation.

\begin{equation}\label{U_t}
    U_t = \sum_i e^{i\omega_i t} \ket{i}\bra{i},
\end{equation}
And demand that the interaction Hamiltonian, $H_{U_t}$ in the interaction frame of $U$, be time-independent. 

\begin{figure}
    \centering
    \includegraphics[width=0.75\linewidth]{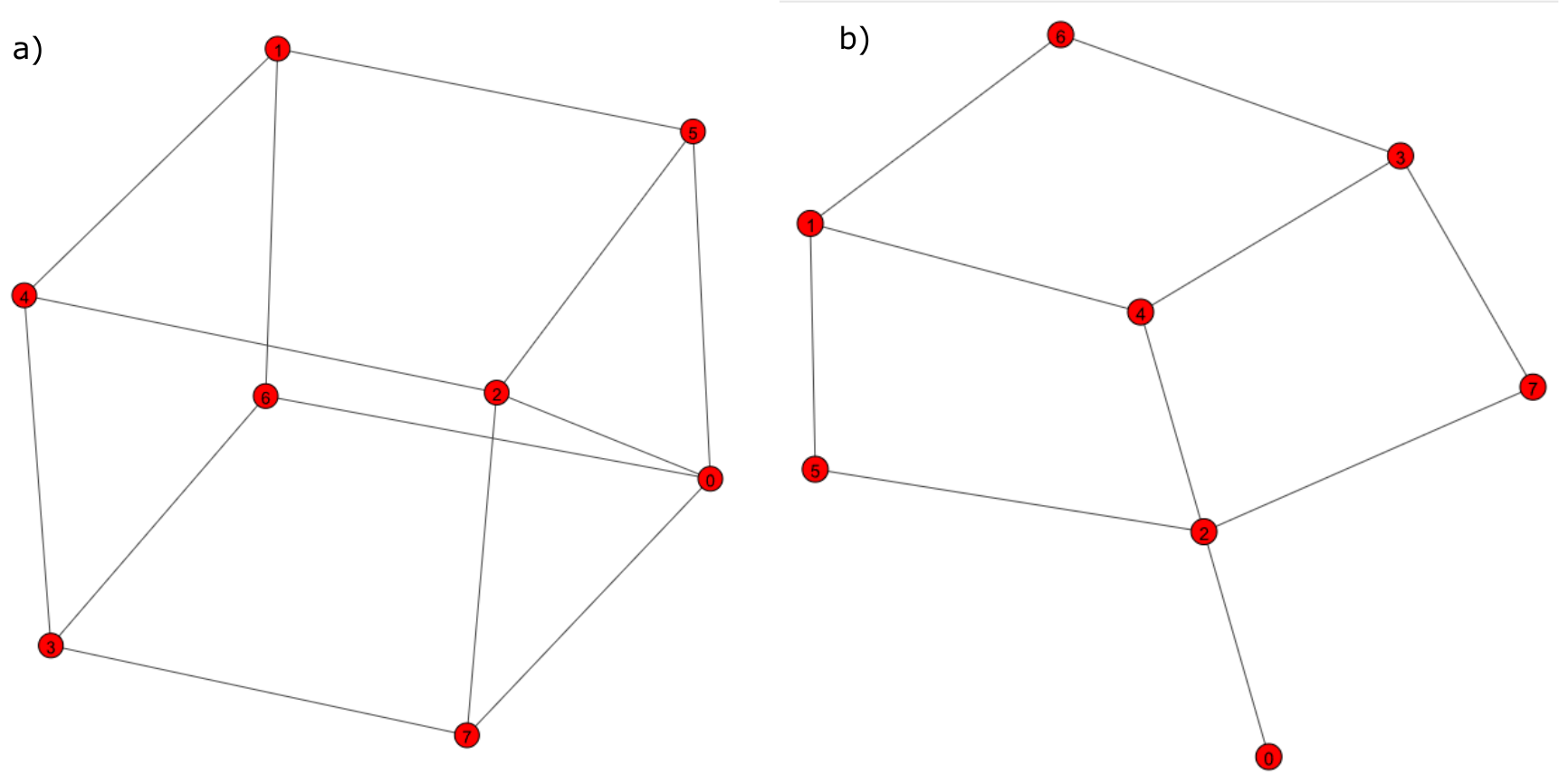}
    \caption{Coupling graphs for \Yb{} ion interacting with light and microwave radiation}
    \label{fig:graphApproach}
    \textbf{a)} All allowed \Shalf to \Phalf transitions and microwave coupling of the clock states.
    We shall define this as the system A. \\
    \textbf{b)} All couplings in \textbf{a)} except \Shalf F=0 to \Phalf F$^\prime$=1.
    We shall define this as the system B.
\end{figure}

To solve for $H_{U_t}$, we use the graph theoretic approach in Ref. \cite{Einwohner1976AnalyticalApproximation}.
We first start by drawing graphs that describe the system and the couplings as shown in Fig. \ref{fig:graphApproach} a) and b).
In Fig. \ref{fig:graphApproach} a), we consider all the couplings defined in Eq. (\ref{eq:CH4totalH}) and in Fig. \ref{fig:graphApproach} b) where we have left out the couplings of F = 0 $\rightarrow$ F $^\prime$ = 1 or levels coupled through light at frequency $\omega_{\mathrm{cool}}$.
Here, the vertices of the graph are the energy levels, and only the energy levels coupled by light are connected through the edges.
Once the edges are drawn, we assign weights to the edges connecting the $i^{th}$ and $j^{th}$ vertices such that the weight of each edge is the `effective detuning', i.e., the coefficient in the exponential in the $(i,j)^{th}$ term (The order of $i$ and $j$ matters, so this is not precisely the weight) in the matrix representation of $H_{\mathrm{rot}}$ from Eq. (\ref{eq:CH4Hrot}).
Now we identify all the closed loops in the graph and if the sum of the `effective detunings' in each closed loop is zero then the system is solvable, i.e. the time dependence of the Hamiltonian can be removed.
Note that while summing the detunings in a given closed loop, the vertices need to be iterated in order, and the calculation of the effective detunings should take this order into account.

For the graph in Fig. \ref{fig:graphApproach} b) (system B), the effective detunings on all closed loops add up to zero; however, for the graph in Fig. \ref{fig:graphApproach} a) (system A) the effective detunings do not add up to zero for all closed loops.
This means that for all values of $\Delta_{\mathrm{micro}},\Delta_{\mathrm{opt}},\Delta_{\mathrm{det}}$, the time dependence in system B can be removed; however, this is not true for the case of system A where the time dependence can be removed only for specific values of $\omega_{\mathrm{cool}}$.
To assign values to all $\omega_i$s in Eq. \ref{U_t}, we proceed with the graph for system B in the following order.

\begin{itemize}
    \item set $\omega_0 = 0$
    \item solve for $\omega_2 $
    \item solve for $\omega_4$, $\omega_5$ and $\omega_7$
    \item solve for $\omega_1$ and $\omega_3$
    \item solve for $\omega_6$
\end{itemize}

Plugging the values of $\omega_i$ from above into $H_{U_t}$ and defining $\delta \coloneq \Delta_{\mathrm{cool}} - \Delta_{\mathrm{micro}} - \Delta_{\mathrm{opt}}$, we find that $H_{U_t}$ is time independent except for the terms where we had $\omega_{cool}$ where the time dependence is now at a smaller frequency of $\delta$.
Since in this work our couplings are set such that $\delta = 0$ the Hamiltonian is fully time independent.
We find that the simplified Hamiltonian takes the form described in Eq. \ref{eq:timeIndependentHamiltonian}.

\begin{equation}
    \begin{split}
        H_{int}^{\prime} &= UHU^{\dagger} + i\hbar \dot U U^{\dagger} \\ 
        & = \sum_i \Delta_i \kb{i}{i} + \sum_{i,j} \frac{\Omega_{ij}}{2} \left( \kb{i}{j} + \kb{j}{i} \right).         
\end{split}    
\end{equation}

with

\begin{equation}\label{eq:CH4H_Udiagonals}
    \begin{split}
        \Delta_1 &= -\Delta_{\mathrm{mic}} - \omega_{\mathrm{zm}}^\mathrm{S} \\
        \Delta_2 &= -\Delta_{\mathrm{mic}} \\
        \Delta_3 &= -\Delta_{\mathrm{mic}} + \omega_{\mathrm{zm}}^\mathrm{S} \\
        \Delta_4 &= -\Delta_{\mathrm{det}} - \Delta_{\mathrm{mic}} \\
        \Delta_5 &= -\Delta_{\mathrm{mic}} - \Delta_{\mathrm{opt}} - \omega_{\mathrm{zm}}^\mathrm{P} \\
        \Delta_6 &= -\Delta_{\mathrm{mic}} - \Delta_{\mathrm{opt}} \\
        \Delta_7 &= -\Delta_{\mathrm{mic}} - \Delta_{\mathrm{opt}} + \omega_{\mathrm{zm}}^\mathrm{P}.
    \end{split}
\end{equation}

Since \Shalf to \Phalf is a dipole transition, \Phalf can decay back to \Shalf through spontaneous emission.
The spontaneous emission results from the coupling of the atom (ion) to the vacuum modes of radiation and can be accounted for through a master equation approach as shown in \cite{Agarwal1970Master-EquationEmission}.
Instead of solving the Schrodinger's equation for the system, we now have to solve for the Lindblad (or Gorini-Kossakowski-Sudarshan-Lindblad) master equation.

\begin{equation}
    \dot{\rho}(t)=-\frac{i}{\hbar}[H_{U_t}, \rho(t)]+\sum_n \frac{1}{2}\left[2 C_n \rho(t) C_n^{+}-\rho(t) C_n^{+} C_n-C_n^{+} C_n \rho(t)\right].
\end{equation}
Where $\rho$ is the density matrix, and $C_n = \sqrt{\Gamma_n} A_n$ are the collapse operators, with $A_n$ being the operators that describe the action the environment induces on the system.
The collapse operators that account for the spontaneous emission of the population in \Phalf into \Shalf are 

\begin{equation}
\begin{aligned}
    C_{14} = \sqrt{\frac{\Gamma_{\mathrm{P}}}{3}} \kb{1}{4}, 
    C_{24} = \sqrt{\frac{\Gamma_{\mathrm{P}}}{3}} \kb{2}{4}, 
    C_{34} = \sqrt{\frac{\Gamma_{\mathrm{P}}}{3}} \kb{3}{4},\\
    C_{05} = \sqrt{\frac{\Gamma_{\mathrm{P}}}{3}} \kb{0}{5},
    C_{15} = \sqrt{\frac{\Gamma_{\mathrm{P}}}{3}} \kb{1}{5},
    C_{25} = \sqrt{\frac{\Gamma_{\mathrm{P}}}{3}} \kb{2}{5},\\ 
    C_{06} = \sqrt{\frac{\Gamma_{\mathrm{P}}}{3}} \kb{0}{6},
    C_{16} = \sqrt{\frac{\Gamma_{\mathrm{P}}}{3}} \kb{1}{6}, 
    C_{36} = \sqrt{\frac{\Gamma_{\mathrm{P}}}{3}} \kb{3}{6},\\
    C_{07} = \sqrt{\frac{\Gamma_{\mathrm{P}}}{3}} \kb{0}{7},
    C_{27} = \sqrt{\frac{\Gamma_{\mathrm{P}}}{3}} \kb{2}{7}, 
    C_{37} = \sqrt{\frac{\Gamma_{\mathrm{P}}}{3}} \kb{3}{7}.    
\end{aligned}
\end{equation}

\subsection{Rabi Frequencies}
The rabi frequencies for a simple 2-level system are set according to the formula: 

\begin{equation*}
    \frac{I}{I_{\mathrm{sat}}}=\frac{2\Omega^2}{\Gamma^{2}},
\end{equation*}

where  $I$ is the intensity of the laser, $I_{\mathrm{sat}}$ is the saturation intensity, and $\Gamma$ is the spontaneous emission rate of the transition. In our case, we are interested in finding the rabi frequency pertaining to a specific transition i.e  
\begin{equation*}
\begin{aligned}
    \Omega_{F,m_{F},F',m_{F'}  } = \frac{ \mel{ F,m_{F}}{d \cdot E}{F',m_{F'} } }{\hbar}
\end{aligned}
\end{equation*}
Applying the Wigner-Eckart theorem, we get

\begin{equation*}
\begin{split}
\mel{ F,m_{F}}{d_q}{F',m_{F'} } = \braket{ F,m_{F}}{F',m_{F'},1,q } \mel{F}{|d_q|}{F'} 
\end{split}
\end{equation*}
Now, the reduced matrix element can be further broken down as 

\begin{equation*}
    \begin{split}
         \mel{F}{|d|}{F'}  =& \mel{J I F}{|d|}{J' I F'} \\
    =& \mel{J}{|d|}{J'} (-1)^{F'+J+1+I}  \\
     & \sqrt{ (2F' + 1)  (2J + 1)}\Sixj{J}{J'}{1}{F'}{F}{I}
    \end{split} 
\end{equation*}

The reduced matrix element between the J levels can simply be calculated from the decay rate of the excited state using Fermi's golden rule as follows:

\begin{equation*}
    \Gamma_{Jg Je} = \frac{\omega_0^3}{3 \pi \epsilon_0 \hbar c^3} \frac{2J_g + 1}{2J_e + 1} |\mel{J_g}{|d|}{J_e}|^2
\end{equation*}

Since it is commonly used in \Yb literature, we introduce the saturation intensity as defined for the $^2$S$_{1/2}$ to $^2$P$_{1/2}$ ignoring the internal structure:

\begin{equation*}
    I_{\mathrm{sat}} = \frac{\pi \Gamma c h}{3 \lambda^{3}}
\end{equation*}
 Combining these equations along with  $|E| = \sqrt{ I / 2 c \epsilon_0 }$ we get

 \begin{equation*}
 \begin{split}
     \Omega^2_{ F,m_{F},F',m_{F'} } &= \frac{I}{I_{\mathrm{sat}}} \frac{\Gamma_{ Jg Je}^2 }{ 2 } |\braket{ F,m_{F}}{F',m_{F',1,q} }|^2 \times \\
     & (2F' + 1)  (2J + 1)   \Sixj{J}{J'}{1}{F'}{F}{I}^2 \times 
       \frac{2J_e + 1}{2J_g + 1} 
 \end{split}
\end{equation*}

In the case of \Yb, for all the allowed transitions between  $^2$S$_{1/2}$ and $^2$P$_{1/2}$, the second line of the above expression evaluates to 1/3, leaving us with a particularly simple expression for the rabi frequency 
\begin{equation}
    \Omega^2 = \frac{I}{I_{\mathrm{sat}}} \frac{\Gamma^2}{6}
\end{equation}



\end{document}